**Significant dependence of the efficiency of energy-saving thermochromic $VO_2$ on slight changes of its properties in the visible due to strain and/or vacancies**

Jiri Houska

*Department of Physics and NTIS - European Centre of Excellence, University of West Bohemia, Univerzitni 8, 30614 Plzen, Czech Republic, email jhouska@kfy.zcu.cz*

**Abstract**

There are worldwide efforts to maximize the energy saving achieved by $VO_2$-based thermochromic coatings. In particular, there are very different values of the modulation of integral solar energy transmittance, reported by various laboratories on various templates even for seemingly very similar coatings. A detailed analysis reveals that this is largely due to the combination of the intentional and well understood transmittance modulation in the infrared (always beneficial) with not yet understood slight transmittance modulation in the visible (sometimes beneficial, sometimes harmful, and always multiplied by strong solar irradiance). Ab-initio calculations are used to examine the hypothesis that the transmittance modulation in the visible can be controlled by lattice strain and/or slightly off-stoichiometric [O]/[V] ratio. The presented phenomenon opens a pathway which may lead, in a case of reproducible preparation of correctly altered $VO_2$, to significantly enhanced energy saving.

**1. Introduction**

Vanadium dioxide is presently the most investigated thermochromic material. Its two key polymorphs include the low-temperature semiconductive monoclinic (distorted rutile) phase $VO_2$(M1), and the high-temperature metallic tetragonal (perfect rutile) phase $VO_2$(R) [1]. Thus, heating of $VO_2$ above its transition temperature ($T_{tr}$) leads to closing of the narrow band gap, to orders of magnitude higher concentration of free charge carriers, and to a strong modulation of all properties which the charge carriers are responsible for. This includes enhanced extinction coefficient especially in the infrared, enhanced electrical conductivity or enhanced electronic part of thermal conductivity. The modulation of functional properties can be used in many applications, including, probably in the first place, energy-saving smart windows which transmit infrared radiation in the cold state but reflect a significant part of infrared radiation in the hot state. See the recent reviews [2-8].

The research in this field [9-38] is focused on two groups of challenges. First, we need the ability to prepare the desired material at all, ideally by a technique (such as magnetron sputtering) which is really used in large industrial glass coaters, and under industry-friendly conditions (low preparation temperature, no substrate bias voltage). This is far from trivial: $VO_2$ is only one of many stoichiometries, $VO_2$(M1⇔R) is only one of many $VO_2$ polymorphs, and the preparation technique should include the possibility to shift $T_{tr}$ to the room temperature (most often by doping of $VO_2$ by W) at preserved strength of the thermochromic transition. While our lab is involved in these efforts [33-35], the scope of the present paper is different.

Second, we need to maximize the quantified performance of thermochromic coatings in terms of integral luminous transmittance ($T_{lum}$) and modulation of integral solar energy transmittance ($\Delta T_{sol}$). See Eqs. 1 and 2, respectively, where $T(\lambda,T_m)$ is a spectral transmittance at a measurement temperature $T_m$, $\varphi_{lum}(\lambda)$ is the luminous sensitivity of human eye and $\varphi_{sol}(\lambda)$ is the solar irradiance. If we let alone the trade-off between $T_{lum}$ and $\Delta T_{sol}$ resulting from varied thickness of the thermochromic layer, there are ideas how to optimize both quantities in parallel. Examples include second-order antireflection layers [36], localized surface plasmon resonance in the high-temperature metallic state [37], or once again doping [38].

$$T_{lum} = \int_{380}^{780} \varphi_{lum}(\lambda)\varphi_{sol}(\lambda)T(\lambda,T_m)d\lambda \Big/ \int_{380}^{780} \varphi_{lum}(\lambda)\varphi_{sol}(\lambda)d\lambda \qquad (1)$$

$$T_{sol} = \int_{300}^{2500} \varphi_{sol}(\lambda)T(\lambda,T_m)d\lambda \Big/ \int_{300}^{2500} \varphi_{sol}(\lambda)d\lambda \quad \text{and} \quad \Delta T_{sol} = T_{sol}(T_m < T_{tr}) - T_{sol}(T_m > T_{tr}) \qquad (2)$$

The aim of the present work is to elaborate another idea, briefly mentioned in the review published by the author in 2022 [2]. It builds on the fact that the overall transmittance modulation includes not only the aforementioned well known strong modulation in the infrared but also much less investigated slight modulation in the visible. While the former is multiplied by only moderate solar irradiance when calculating $\Delta T_{sol}$, the latter is multiplied by strong solar irradiance when calculating $\Delta T_{sol}$. Because higher modulation of $T(\lambda,T_m)$ is multiplied by lower $\varphi_{sol}(\lambda)$ and vice versa, both these wavelength ranges can constitute comparable (in absolute value) contributions to the total $\Delta T_{sol}$. Furthermore, while the contribution of the $T(\lambda,T_m)$ modulation in the infrared always has the desired sign (increasing $\Delta T_{sol}$), the modulation of $T(\lambda,T_m)$ in the visible can have both signs (increasing [39-41] or decreasing [42-44] $\Delta T_{sol}$). The impression given by the literature is that while couple of authors [39-41] achieved and published visible-range modulation of the desired sign and the resulting high $\Delta T_{sol}$, these findings remain rather empirical than knowledge-based. Thus, there is an urgent need to investigate this phenomenon, constituting a large energy-saving potential.

A case can be made that the reason behind this phenomenon may be related to crystalline templates such as $Cr_2O_3$ [39] (it is not easy to be more specific without knowing the crystal orientation) and/or to the exact composition of $VO_{2\pm x}$ [2]. Thus, the present paper, far from exhausting this direction of research, is focused on (i) detailed presentation of the aforementioned phenomenon using our experimental data, and on (ii) using ab-initio calculations to clarify whether the modulation of $T(\lambda,T_m)$ in the visible can be significantly altered in the desired direction by lattice strain and/or slightly off-stoichiometric [O]/[V] ratio.

## 2. Methods

*2.1 Experiment*

The phenomenon studied is illustrated by detailed characteristics of two $VO_2$ coatings prepared in our lab [45,46] (papers which report their basic characteristics, but do not use and discuss them in the present context). One of these coatings (prepared on amorphous glass) exhibits negative transmittance modulation in the visible, while the other one (prepared on crystalline Si) exhibits positive modulation of measured extinction coefficient and calculated transmittance in the visible. Both coatings were prepared by reactive high-power impulse magnetron sputtering (HiPIMS; ATC 2200-V AJA International Inc. system with a TruPlasma Highpulse 4002 Hüttinger Elektronik power supply) of V target in Ar+$O_2$ plasma. The substrates were at a floating potential. The duty cycle of the voltage pulses was 1% and the voltage pulse duration was 50 µs (frequency 200 Hz; glass substrate) or 80 µs (frequency 125 Hz; Si substrate). The deposition-averaged power density was around 13 Wcm$^{-2}$, the pulse-averaged power density was around 1.3 kWcm$^{-2}$ and the peak power density was 5 kWcm$^{-2}$. The Ar partial pressure was 1 Pa at a fixed Ar flow rate of 60 sccm. The specifics of this technique, delivering of high energy and momentum into the growing films by highly ionized fluxes of film-forming atoms, allowed us to prepare the coatings at a floating potential and a very low growth temperature of 250 °C (Si substrate) - 300 °C (glass substrate).

The elemental composition close to $VO_2$ was ensured by pulsed $O_2$ flow control: the flux of $O_2$ was not constant, but varied between 0 and 2 sccm by a logical controller which compares the actual and preset value of selected control quantity (in the present case $O_2$ partial pressure using a preset critical value of 15 mPa, in other cases target current). According to the phase analysis (Raman spectra and XRD patterns), the composition of at least the dominant crystalline (⇔ potentially thermochromic) part of both samples is close to $VO_2$. The sample which does not exhibit the desired modulation in the visible may be slightly O-rich (there is a tiny peak of $V_4O_9$) [45]. The sample which does exhibit the desired modulation in the visible may be slightly V-rich (increasing $k(\lambda)$ in the infrared even for the semiconducting phase, shown also below) [46].

The optical constants $n(\lambda)$ and $k(\lambda)$ were measured by variable-angle spectroscopic ellipsometry using the J.A.Woollam VASE instrument equipped with the Instec heat/cool stage. The measurements were performed in reflection under angles 55, 60 and 65° (glass substrate) or 65, 70 and 75° (Si substrate), at temperatures $T_m = 25$ °C and 100 °C (safely below and above $T_{tr}$ of undoped $VO_2$, respectively). The raw data were processed by an optical model implemented in the WVASE software and containing the substrate, $VO_2$ layer represented by a combination of one Cody-Lorentz oscillator with several Lorentz oscillators, and a surface roughness layer. The coating transmittance, resulting from measured (this subsection) or calculated (next subsection) optical constants, was calculated by the WVASE software which utilizes the transfer matrix formalism.

See our previous papers for further information such as time dependencies of voltage, current and $O_2$ partial pressure [33], energy-resolved mass spectroscopy [33], using deep oscillation magnetron sputtering instead of conventional HiPIMS [34], shift of $T_{tr}$ toward the room temperature by doping [35], enhancement of $T_{lum}$ and $\Delta T_{sol}$ by antireflection layers [36] or the effect of key process parameters on optical properties and electrical resistivity [33,45].

*2.2 Modelling*

The experimental data are complemented by ab-initio calculations utilizing Hubbard-corrected density-functional theory (DFT+$U$ [47]) as implemented in the Quantum Espresso (QE) package [48]. Note that the aim of these calculations is orthogonal to those reported for unstrained stoichiometric $VO_2$ previously [49-56]: neither to further develop the exceptionally challenging methodology of calculations of electronic structures of these strongly correlated materials, nor to reveal the effect of doping in the metal sublattice, but to provide information about possible effects of lattice strain and off-stoichiometric [O]/[V] ratio. The information which can be and is sought are not exact values of optical constants at exact $\lambda$, but whether it is realistically possible to achieve a quantitatively significant effect in the desired direction.

The technicalities shared by all steps of each calculation include expanding the valence electron wavefunction in a plane-wave basis set, treating of exchange and correlation effects using the PBE functional [57], Marzari-Vanderbilt smearing [58] of the occupations of electronic states around the Fermi level, and the $U$ value for V 3d electrons in DFT+$U$, employed to compensate the usual band gap underestimation of DFT, of 3.5 eV (following previously used 3.4 eV [53] - 3.5 eV [56]). The calculations were not spin polarized (no local magnetic moments were observed in $VO_2$(M1), see e.g. the discussion in [52]).

The calculation procedure at a given composition and given $b/a$ (next paragraph) was the following. First, about ten different isotropic deformations have been applied, followed by relaxation of atomic positions and calculation of the total energy. Second, the lowest-energy volume has been identified by fitting the Birch equation of state, and the atomic positions have been relaxed also at this volume. In these first two steps the atom cores and inner electron shells were represented by ultrasoft Vanderbilt pseudopotentials [59], the wavefunction cutoff was 25 Ry, the k-point grid was 6×6×6 for the 12-atom cell and the smearing around the Fermi level was 0.5 eV. Third, the electronic structure has been recalculated using norm-conserving Troullier-Martins pseudopotentials [60], wavefunction cutoff of 60 Ry, k-point grid 12×12×12 for the 12-atom cell and smearing around the Fermi level of 0.1 eV. Fourth, the electronic structure has been used to calculate the dielectric tensor (spectral dependencies of real and imaginary part of the relative permittivity $\varepsilon_x(\lambda)$, $\varepsilon_y(\lambda)$ and $\varepsilon_z(\lambda)$) by the postprocessing code which is a part of the QE package. Fifth, the isotropic refractive index ($n(\lambda)$) and extinction coefficient ($k(\lambda)$) have been predicted by assuming equal volume fractions of crystals characterized by $\varepsilon_x$, $\varepsilon_y$ and $\varepsilon_z$, and using Bruggemann effective medium approximation with a depolarization factor of 1/3. Sixth, transmittance has been calculated using the software VASE for 1 mm thick glass coated by (i) 50 nm thick layer of the investigated $VO_2$ and by (ii) state-of-the-art [36] three-layer coating combining 50 nm thick middle layer $VO_2$ with second-order antireflection layers made of 180 nm of $ZrO_2$ below and above it.

For each strain and each composition, the calculation has been performed for monoclinic $VO_2$(M1) (12-atom primitive cell characterized in the unstrained state by $b/a$ = 0.786523, $c/a$ = 0.935922 and $\cos\beta$ = -0.539447) as well as for tetragonal $VO_2$(R) (6-atom primitive cell characterized in the unstrained state by $c/a$ = 0.626048; may be converted to 12-atom cell resembling $VO_2$(M1) and characterized by $b/a$ = 0.798660, $c/a$ = 0.942262 and $\cos\beta$ = -0.530638). The role of strain has been investigated by changing $b/a$ from 90% to 110% of the preferred (unstrained) value at fixed $c/a$ and $\cos\beta$ (the simplest out of many possibilities, but sufficient to fulfil the present aim), and the role of stoichiometry has been investigated by moving from 6-atom or 12-atom primitive cells to 48-atom supercells with up to 2 oxygen vacancies (the experimental data below indicate desired modulation in the visible in parallel to a relatively strong role of free charge carriers in the infrared, so understoichiometry in O has been considered more promising than overstoichiometry in O).

## 3. Results and discussion

The optical constants of both investigated polymorphs of $VO_2$ prepared on amorphous glass are shown in Fig. 1a. The best known consequence of the thermochromic transition from semiconductive $VO_2$(M1) to metallic $VO_2$(R) is significantly enhanced $k(\lambda)$ in the infrared. However, the information which is even more important for the present work is slightly lower $k(\lambda)$ of $VO_2$(R) in the UV range and in most of the visible range up to ≈645 nm (collectively representing ≈70% of solar energy in UV+VIS and ≈40% of total solar energy). The optical constants of the same polymorphs of $VO_2$ prepared on crystalline Si are shown in Fig. 1b. While the overall qualitative impression is similar to that given by Fig. 1a, there is an important quantitative difference: $k(\lambda)$ of $VO_2(R)$ is higher than that of $VO_2(M1)$ not only in the infrared but also in whole visible range. It is also worth to point out the difference between both figures in terms of $k(\lambda)$ of $VO_2(M1)$ in the infrared. In Fig. 1a this quantity decreases with $\lambda$, indicating negligible contribution of free charge carriers (negligible Drude term of permittivity), while in Fig. 1b this quantity increases with $\lambda$, indicating much stronger role of free charge carriers (arguably due to a slightly more metallic composition $VO_{2-x}$). How closely related is the latter observation to the desired sign of visible-range modulation is an important open question.

The spectral transmittance, calculated using the optical constants from Figs. 1a,b, is shown in Figs. 1c,d (the agreement of transmittance predicted using this methodology and measured transmittance has been confirmed previously [2]). Although the relationship between $k(\lambda)$ and $T(\lambda)$ is not straightforward (role of $n(\lambda)$ and of the interference), in the present case the trends are the same. The negative visible-range modulation of $k(\lambda)$ in Fig. 1a is largely responsible for the negative visible-range modulation of $T(\lambda)$ in Fig. 1c, while the positive visible-range modulation of $k(\lambda)$ in Fig. 1b is largely responsible for the positive visible-range modulation of $T(\lambda)$ in Fig. 1d. These trends are valid for both coating designs considered: $VO_2$(50 nm) layers on glass as well as $ZrO_2$(180 nm) - $VO_2$(50 nm) - $ZrO_2$(180 nm) trilayers on glass. The figure also indicates, beyond the scope of the present paper, the advantage of second-order antireflection layers (180 nm of $ZrO_2$ with $n_{550} = 2.15$; figure S1 in the Supplementary material) in terms of not only $T_{lum}$ but also $\Delta T_{sol}$. Second-order maxima in the visible lead to first-order maxima at approximately 3× longer $\lambda$ in the infrared, and the enhanced $T(\lambda)$ values in the infrared lead to enhanced $T(\lambda)$ modulation in the infrared [36].

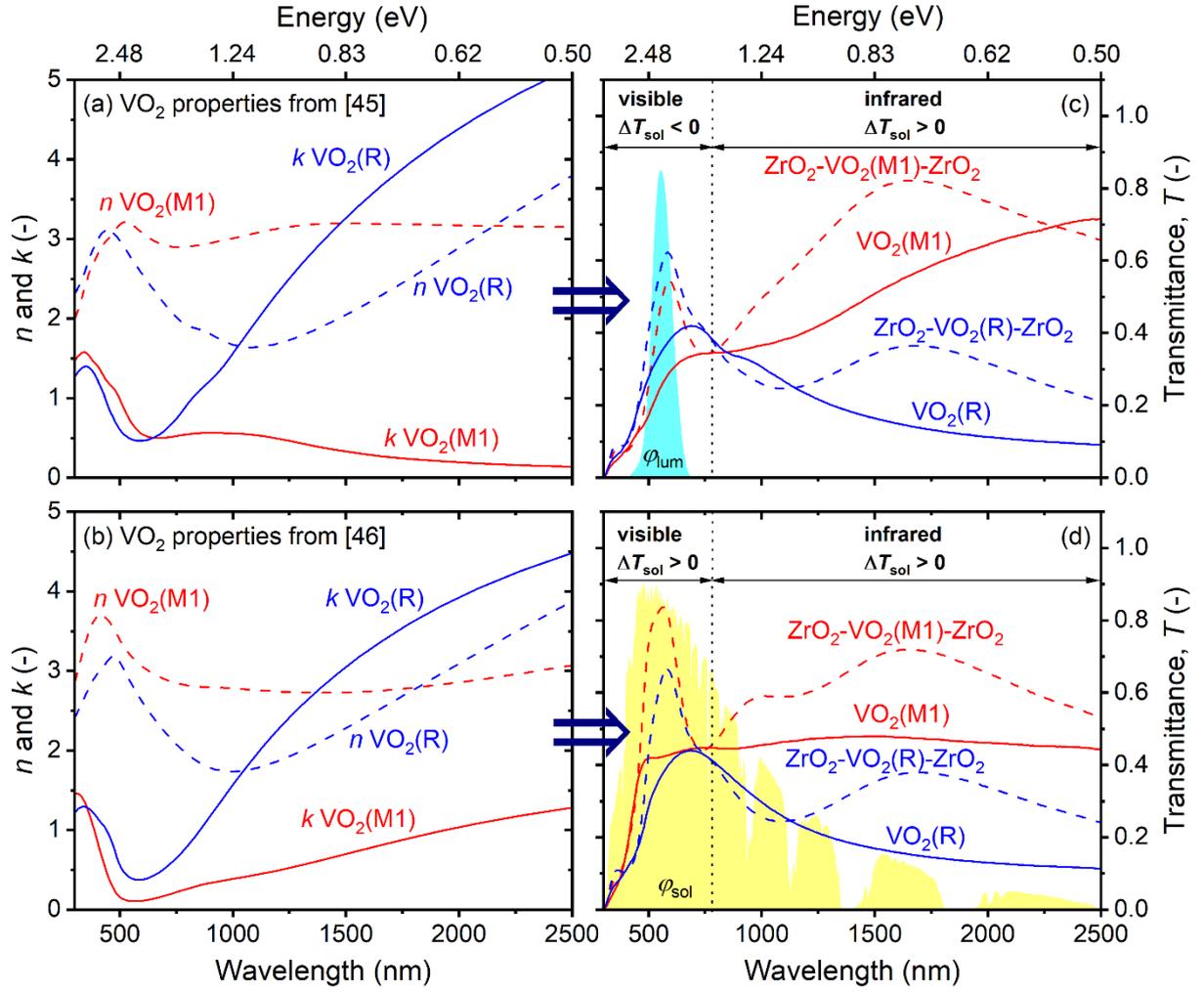

**Figure 1**: Measured refractive index, $n$, of $VO_2$ (dashed lines in panels a[45] and b[46]), measured extinction coefficient, $k$, of $VO_2$ (full lines in panels a[45] and b[46]), calculated transmittance, $T$, of $VO_2$(50 nm) on glass (full lines in panels c,d) and calculated transmittance of $ZrO_2$(180 nm) - $VO_2$(50 nm) - $ZrO_2$(180 nm) trilayers on glass (dashed lines in panels c,d). Red lines are for the low-temperature semiconductive polymorph $VO_2$(M1) and blue lines are for the high-temperature metallic polymorph $VO_2$(R). The shaded areas in panels c and d (arbitrary units without any vertical axis) correspond to the luminous sensitivity of human eye $\varphi_{lum}$ and spectral solar irradiance $\varphi_{sol}$, respectively. Note the qualitative difference between visible-range modulation of $k(\lambda)$ in panels a,c (leading to opposite signs of the integral contributions of visible-range and infrared-range wavelengths to $\Delta T_{sol}$; amorphous glass template) and in panels b,d (the same signs of both contributions; crystalline Si template).

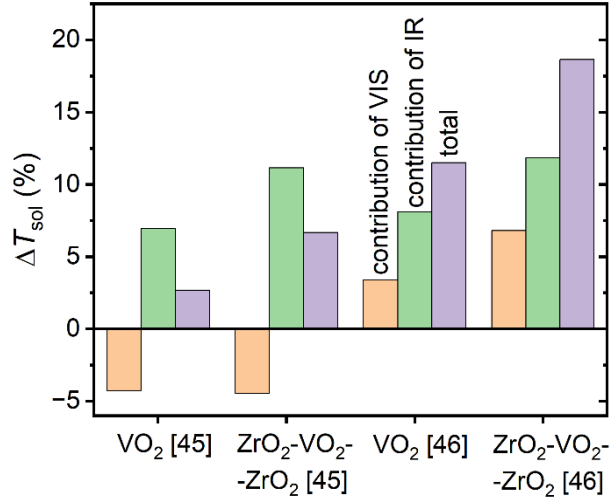

**Figure 2**: Calculated contribution of visible-range light (orange columns) and infrared light (green columns) to the total modulation of integral solar energy transmittance, $\Delta T_{sol}$ (violet columns) for $VO_2$(50 nm) on glass and $ZrO_2$(180 nm) - $VO_2$(50 nm) - $ZrO_2$(180 nm) trilayers on glass. The measured $VO_2$ properties used to calculate $\Delta T_{sol}$ correspond to Fig. 1a (first and second set of columns) and Fig. 1b (third and fourth set of columns).

The qualitative impression given by Fig. 1 is quantified in Fig. 2. The measure of energy saving, $\Delta T_{sol}$, is shown using triplets of columns: contribution of predominantly visible wavelengths from 300 to 780 nm (the importance of near UV is limited by transmittance of the glass substrate), contribution of infrared wavelengths from 780 to 2500 nm, and the total $\Delta T_{sol}$ value given by the sum of the previous two. On the one hand, in the case of $VO_2$ properties shown in Fig. 1a, the beneficial positive modulation in the infrared is to a large extent wasted because of the negative modulation in the visible: $\Delta T_{sol}$ = -4.3% (VIS) + 7.0% (IR) = 2.7% for $VO_2$(50 nm), and $\Delta T_{sol}$ = -4.5% (VIS) + 11.2% (IR) = 6.7% for $ZrO_2$(180 nm) - $VO_2$(50 nm) - $ZrO_2$(180 nm). On the other hand, in the case of $VO_2$ properties shown in Fig. 1b, the beneficial positive modulation in the infrared is supported by also positive modulation in the visible: $\Delta T_{sol}$ = 3.4% (VIS) + 8.1% (IR) = 11.5% for $VO_2$(50 nm), and $\Delta T_{sol}$ = 6.8% (VIS) + 11.9% (IR) = 18.7% for $ZrO_2$(180 nm) - $VO_2$(50 nm) - $ZrO_2$(180 nm).

Indeed, the integral transmittances once again confirm that the dramatical differences resulting from considering $VO_2$ properties from different sources take place only in the visible. It is also worth to add that the advantage of visible-range modulation of $VO_2$ properties obtained in [46] is given not only by its direct consequences (positive contribution of VIS to $\Delta T_{sol}$ in the third set of columns in Fig. 2), but also by the possibility to tailor the second-order antireflection layer in that way that its functionality is higher in the low-temperature state of the coating (2× higher contribution of VIS to $\Delta T_{sol}$ in the fourth than in the third set of columns in Fig. 2). On the contrary, the role of infrared wavelengths is almost independent of the source of $VO_2$ properties, for $VO_2$(50 nm) (contribution of IR to $\Delta T_{sol}$ in the first and third set of columns in Fig. 2 of ≈7-8%) as well as for $ZrO_2$(180 nm) - $VO_2$(50 nm) - $ZrO_2$(180 nm) (contribution of IR to $\Delta T_{sol}$ in the second and fourth set of columns in Fig. 2 of ≈11-12%, enhanced by the aforementioned first-order interference maximum in the infrared).

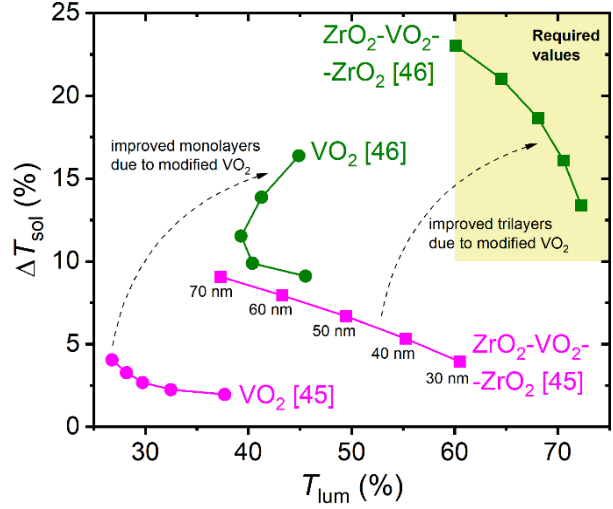

**Figure 3**: Calculated average luminous transmittance, $T_{lum}$, and modulation of integral solar energy transmittance, $\Delta T_{sol}$, for $VO_2$ monolayers on glass (balls) and $ZrO_2$(180 nm) - $VO_2$ - $ZrO_2$(180 nm) trilayers on glass (squares). The data are shown for $VO_2$ properties corresponding to Fig. 1a (magenta) and Fig. 1b (green), and for $VO_2$ thicknesses of 30 nm to 70 nm with a step of 10 nm. The shaded area represents the required $T_{lum}$ and $\Delta T_{sol}$ values for smart-window applications.

The whole calculated thermochromic performance in terms of not only $\Delta T_{sol}$ but also $T_{lum}$ (averaged over the low- and high-temperature state) can be seen in Fig. 3. The performance is shown not only for the $VO_2$ thickness of 50 nm which is used throughout this paper, but also for other thicknesses from 30 to 70 nm. One of the roles of this figure is to avoid any impression that the altered modulation in the visible can improve $\Delta T_{sol}$ only at a cost of worse $T_{lum}$. In the present case where altered modulation means in the first place lowered visible-range $k(\lambda)$ of $VO_2$(M1) rather than e.g. enhanced visible-range $k(\lambda)$ of $VO_2$(R), the figure clearly shows that the transition from $VO_2$ properties in Fig. 1a to those in Fig. 1b improves both $T_{lum}$ and $\Delta T_{sol}$. This is not to be confused with the trade-off between $T_{lum}$ and $\Delta T_{sol}$ resulting from varied $VO_2$ thickness, exhibited by all dependencies in Fig. 3 (except that $T_{lum}$ of the more transparent $VO_2$ monolayer exhibits a non-monotonic dependence on the thickness of this monolayer because of the interference on it). Note that as long as the altered modulation in the visible shifts the whole thickness dependence in the desired top right direction, this trade-off would allow one to change the thickness and improve $\Delta T_{sol}$ at preserved $T_{lum}$ even if the modulation was achieved by enhanced visible-range $k(\lambda)$. The improved properties of $ZrO_2$(180 nm) - $VO_2$ - $ZrO_2$(180 nm) trilayer fulfil the frequent requirement ([2] and Refs. therein) for cost-effective smart-window applications: $T_{lum} \geq 60\%$ and $\Delta T_{sol} \geq 10\%$.

The calculated spectral dependencies of optical constants of both unstrained and strained $VO_2$ are shown in Figs. 4a,b. Although the aforementioned aim is to capture the qualitative effect of strain rather than to predict exact $n(\lambda)$ and $k(\lambda)$, the figures actually reproduce many of the features observed (Fig. 1) in the same wavelength range experimentally. There is (i) $n(\lambda)$ of $VO_2$(M1) of around 3, (ii) drop of minimum values of $n(\lambda)$ after the transition to $VO_2$(R), (iii) $k(\lambda)$ of $VO_2$(M1) ranging from low values in IR to 1.5-2.0 in VIS and near UV and, above all (iv) the main merit of thermochromic $VO_2$: steep increase of $k(\lambda)$ in IR after the transition to $VO_2$(R). The biggest quantitative difference from Fig. 1 is overestimated $k(\lambda)$ of $VO_2$(M1) in the central part of the presented range: reason why Figs. 4c and 5c show ratios rather than values of transmittances. For the comparison of modelling and experiment note also figures S2 and S3 in the Supplementary material.

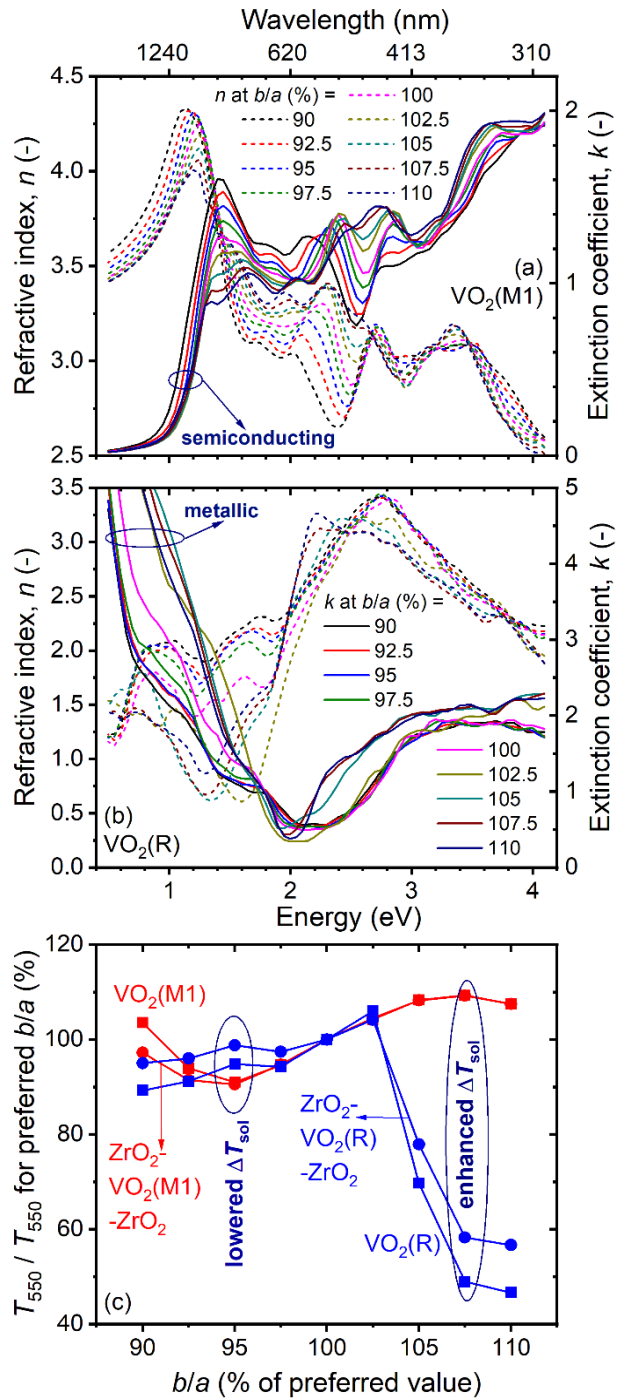

**Figure 4**: Effect of varied lattice constant ratio $b/a$ on calculated refractive index (dashed lines) and extinction coefficient (solid lines) of $VO_2(M1)$ (panel a) and $VO_2(R)$ (panel b). Panel c shows the transmittance at 550 nm, $T_{550}$, calculated using this $n$ and $k$ of both polymorphs for $VO_2$(50 nm) on glass (squares) and $ZrO_2$(180 nm) - $VO_2$(50 nm) - $ZrO_2$(180 nm) trilayers on glass (circles) and shown as a ratio to $T_{550}$ calculated for preferred $b/a$.

Figures 4a,b show that on the one hand, the strain-induced changes preserve the main qualitative feature of the thermochromic transition from $VO_2(M1)$ to $VO_2(R)$: increase of $k(\lambda)$

at all wavelengths above ≈800±100 nm (limit decreasing with $b/a$), i.e. all energies below ≈1.55±0.20 eV (limit increasing with $b/a$). On the other hand, there are complicated strain-induced changes of both $n(\lambda)$ and $k(\lambda)$ in the visible, including different signs of these changes at different wavelengths for the same strain. To put an example for 550 nm = 2.25 eV (middle of the visible range and close to the maximum of $\varphi_{sol}(\lambda)$), increasing $b/a$ of $VO_2$(M1) leads to monotonically increasing $n_{550}$ at mostly decreasing $k_{550}$, and increasing $b/a$ of $VO_2$(R) leads to a convex dependence of both $n_{550}$ and $k_{550}$, at particularly steep increase of $k_{550}$ at $b/a$ enhanced to 105-110% of the preferred value. Although a case can be made that (despite using DFT+$U$ instead of just DFT) 550 nm in simulations does not exactly correspond to 550 nm in the experiment, (i) this is consistent with the aforementioned aim of the present calculations, and (ii) the most significant effect, increase of $k(\lambda)$ of $VO_2$(R) at enhanced $b/a$, is actually observable in Fig. 4b not only at 550 nm but in a wide range of wavelengths.

The presented strain-induced changes of optical constants of both $VO_2$(M1) and $VO_2$(R) directly lead to strain-induced changes of their transmittance. This is shown in Fig. 4c for the same two coating designs as in Figs. 1-3, for brevity only at the aforementioned $\lambda$ = 550 nm = 2.25 eV. The figure confirms the intuition that the effect of lattice strain of $VO_2$ on individual contributions to $\Delta T_{sol}$ may be positive as well as negative. To put a specific example, the effect on the modulation of $T_{550}$ is positive at $b/a$ enhanced to at least 105% and negative at $b/a$ lowered to about 95% of the preferred value, independently of the coating design ($VO_2$(50 nm) or $ZrO_2$(180 nm) - $VO_2$(50 nm) - $ZrO_2$(180 nm)). Quantitatively, the strain-induced changes of $T_{550}$ of both polymorphs at $b/a$ up to 105% are within approximately ±10%, and the changes at $b/a$ = 105-110% are even much more significant. The importance of the "small" changes on the order of 10% is emphasized by their comparison with experimental visible-range contributions to $\Delta T_{sol}$ in Fig. 2, which are on the same order. On the minus side, the positive effect at such enhanced $b/a$ is mostly due to enhanced $k_{550}$ and in turn lowered $T_{550}$ of $VO_2$(R) and only slightly due to enhanced $T_{550}$ of $VO_2$(M1), indicating that the present example of a strain tensor does not constitute the most promising one. It is perfectly likely that other strain tensors used in the future may be even more beneficial and/or easier to realize (intentionally or by chance) using specific crystalline substrates. One reason why the available information is still very limited is rarely reported preferred crystal orientation (partially resulting from the preference of grazing-incidence XRD over Bragg-Brentano XRD in the case of very thin $VO_2$ layers).

The calculated spectral dependencies of optical constant of both $VO_2$ and vacancy-containing $VO_{2-x}$ are shown in Figs. 5a,b. There is a significant qualitative difference between the effect of strain in Figs. 4a,b and the effect of stoichiometry in Figs. 5a,b. While the effect of strain is significant in the visible but does not affect the order of magnitude of low $k(\lambda)$ of $VO_2$(M1) and high $k(\lambda)$ of $VO_2$(R) in the infrared, the effect of oxygen vacancies is the largest in terms of $k(\lambda)$ of $VO_2$(M1) in the infrared.

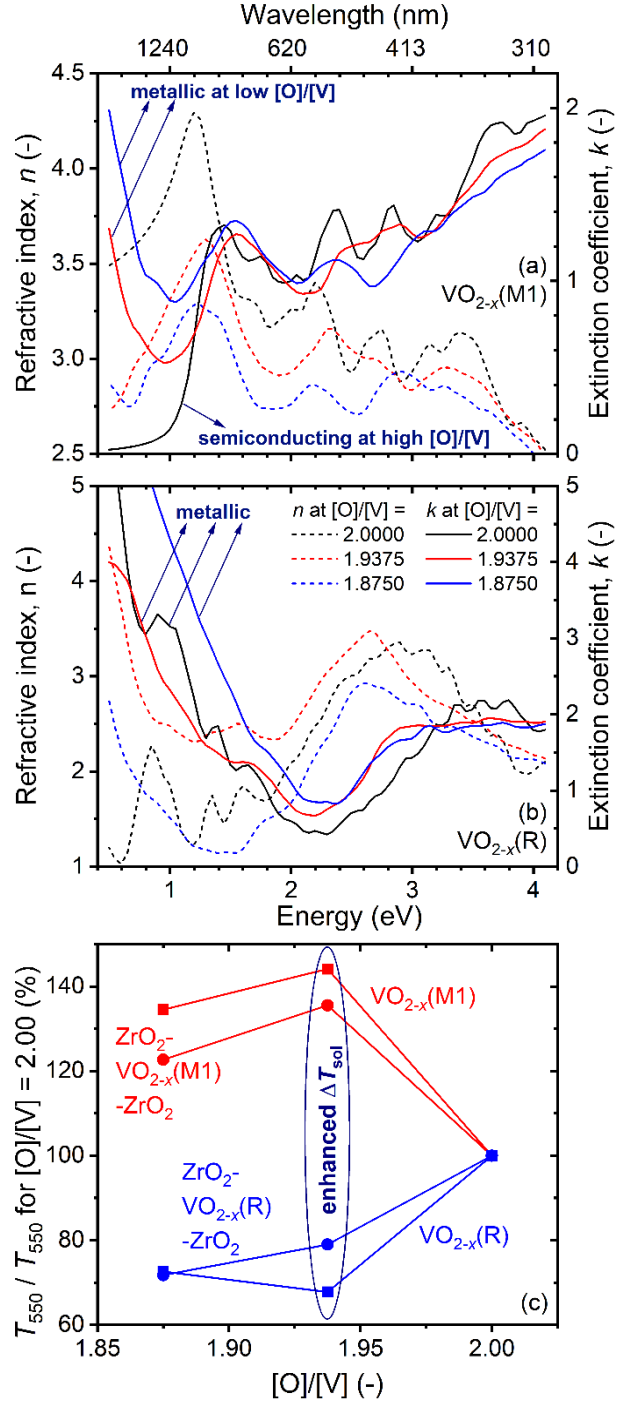

**Figure 5**: Effect of varied composition on calculated refractive index (dashed lines) and extinction coefficient (solid lines) of $VO_{2-x}$(M1) (panel a) and $VO_{2-x}$(R) (panel b). Panel c shows the transmittance at 550 nm, $T_{550}$, calculated using this $n$ and $k$ of both polymorphs for $VO_2$(50 nm) on glass (squares) and $ZrO_2$(180 nm) - $VO_2$(50 nm) - $ZrO_2$(180 nm) trilayers on glass (circles) and shown as a ratio to $T_{550}$ calculated for stoichiometric $VO_2$.

First, $VO_2$(M1) ($V_{16}O_{32}$ simulation cell) once again exhibits, in agreement with the experiment, dispersion characteristic of a semiconductor: decreasing $k(\lambda)$ in the infrared with decreasing energy, at no role of free charge carriers. Second, $VO_{1.9375}$(M1) ($V_{16}O_{31}$ simulation cell), exhibits the opposite: increasing $k(\lambda)$ in the infrared (below $\approx 1$ eV) with decreasing energy, i.e. a usual fingerprint of free charge carriers. In other words, one vacancy per 32 O positions (concentration chosen due to computational costs) is sufficient for a significant Drude term of permittivity, and even lower vacancy concentrations may be thus recommended for future experimental efforts. This is consistent with the fact that $VO_{1.9375} = V_{16}O_{31}$ is actually as far as halfway between $VO_2$ and the next phase in the V-O phase diagram, $V_8O_{15}$. Third, $VO_{1.8750}$(M1) ($V_{16}O_{30}$ simulation cell) confirms this trend: 2× higher vacancy concentration further increases (albeit less than 2×) $k(\lambda)$ in the infrared. On the contrary, the corresponding effect of O vacancies on $VO_{2-x}$(R) is much weaker: this phase is metallic for all compositions investigated, and its $k(\lambda)$ in the infrared is consequently (i) increasing with decreasing energy and (ii) still higher than that of $VO_{2-x}$(M1) (note different vertical scales) regardless the number of vacancies. To put an example of the predicted effect on infrared transmittance, the presence of 1-2 vacancies decreases the transmittance of 50 nm thick $VO_2$(M1) at $\lambda = 2500$ nm from 70% to 36-48% and its modulation after the transition to $VO_2$(R) from 58% to 33-36%.

The effect of vacancies in the visible shown in Figs. 5a,b is much weaker than that in the infrared, and it is (in the present compositional and strain range) on the same order as the effect of lattice strain in the visible shown in Figs. 4a,b. However, this effect is important also here, and for the same reason: multiplication by high $\varphi_{sol}(\lambda)$. If we follow the previous figure and once again put an example for 550 nm = 2.25 eV, it is interesting that the effect of the creation of vacancies is in the same direction as that of the enhanced $b/a$: lower $k_{550}$ of $VO_{2-x}$(M1) and higher $k_{550}$ (increase which is once again observable in a wide range of wavelengths) of $VO_{2-x}$(R).

Again, the presented composition-dependent changes of optical constants of both $VO_2$(M1) and $VO_2$(R) directly lead to composition-dependent changes of their transmittance shown in Fig. 5c. We can observe a different way (incorporation of vacancies instead of increasing $b/a$) how to achieve the desired effect on the modulation of $T_{550}$ and in turn the enhanced energy saving in terms of $\Delta T_{sol}$. Furthermore, contrary to the role of $b/a$, the desired effect is this time achieved not only by lower $T_{550}$ of $VO_2$(R) (improving $\Delta T_{sol}$ at a cost of $T_{lum}$) but equally importantly also by higher $T_{550}$ of $VO_2$(M1) (improving both $\Delta T_{sol}$ and $T_{lum}$ simultaneously).

Thus, the data in Fig. 5 indicate that manipulating the exact composition is equally promising way of achieving the desired visible-range contribution to $\Delta T_{sol}$ as manipulating the lattice deformation. A case can be made that manipulating the composition is even more promising because it can be utilized on any growth template (e.g., any material of antireflection layer below the active thermochromic layer). A synergy of both these options is worth a consideration as well, including the possibility that the former (creation of vacancies) may be induced by the latter (template-dependent lattice strain). This constitutes one possible explanation of the difference between the two samples prepared in a similar way and presented in Fig. 1: no observable role of free charge carriers and harmful modulation in the visible on the amorphous template, fingerprint of free charge carriers and desired modulation in the visible on the crystalline template.

## 4. Conclusions

Modulation of integral solar energy transmittance, $\Delta T_{sol}$, constitutes a measure of energy saving using $VO_2$-based thermochromic coatings. Contributions of visible and infrared range to $\Delta T_{sol}$ can have comparable absolute values. While the contribution of infrared range to $\Delta T_{sol}$ always has the desired sign, the contribution of visible range to $\Delta T_{sol}$ can have both signs. This has been observed even for $VO_2$ layers prepared using very similar deposition protocols except different growth templates.

Ab-initio calculations reveal that the modulation in the visible can be affected by lattice strain as well as by slight changes of the $VO_{2\pm x}$ stoichiometry. For example, desired direction and significant size of this effect has been achieved by increasing $b/a$ to ≥105% of the preferred value, or alternatively by introducing 1-2 O vacancies per 32 O positions. The latter also increases the metallicity of the low-temperature phase $VO_2(M1)$, which is relevant for the explanation of the presented experimental data as well.

Collectively, the results show that the efficiency of energy-saving thermochromic $VO_2$ can be significantly enhanced ($\Delta T_{sol}$ in the presented example from 2.7% to 11.5% or from 6.7% to 18.7% depending on the coating design) by slight changes of its properties in the visible. In order to fully utilize this potential and to avoid the opposite, further experimental as well as theoretical research is needed. On the experimental side this may require investigation of preferred crystal orientation in general and epitaxial growth on various templates in particular, as well as precise compositional measurements. Antireflection layers below $VO_2$ can simultaneously fulfil also the role of crystalline growth template and induce the lattice strain. On the theoretical side this may require extensive calculations considering various strain tensors or even lower vacancy concentrations, as well as further improvement of the exactness of these calculations for strongly correlated $VO_2$. Focus on a combination of both effects (compositional changes induced by a lattice strain) may be relevant on both sides.

**Data availability**

Data will be made available on request.

**Acknowledgment**

This work was supported by the project Quantum materials for applications in sustainable technologies (QM4ST), funded as project No. CZ.02.01.01/00/22_008/0004572 by Programme Johannes Amos Commenius, call Excellent Research. Computational resources were provided by the e-INFRA CZ project (ID:90254), supported by the Ministry of Education, Youth and Sports of the Czech Republic.